# Appropriate transient thermal analysis of an absorber plate in flat-plate solar collectors from beginning to end operational conditions

Rahul Roy*, Balaram Kundu

Department of Mechanical Engineering, Jadavpur University, India

**Abstract**

This work establishes a closed-form analysis for two-dimensional transient heat transfer in an absorber plate based on the thermal wave model when the collector starts to work. The method of separation of variables is employed to determine the exact thermal response during the starting of a solar collector. The thermal field is represented by 2-D thermal contours and 3-D surface plots to provide an accurate prediction compared to the 1-D analysis using the time-lag theory. An influence of Fourier number on the heat flow in the absorber plate is studied for different thermal relaxation times. The present analysis highlights that the magnitude of plate temperature depends significantly on the physical geometry. Differences in two models of heat propagation, viz. classical and thermal wave, in the absorber plate have an alteration in predicted collector performances. The present study also develops another design situation for heat transfer in the absorber plate when the collector shutdowns to the ambient state from its steady condition. Laplace Transform Method (LTM) in combination with the product solution method uses to determine the temperature with the initial condition represented as a steady case. Due to the lagging nature of thermal behavior, a significant difference between the classical and thermal wave models for the propagation of heat was observed, and also this study determines the exact time to be taken to reach the dead state for the complete shutdown of the collector.

**Nomenclature**

| Symbol | Description | Units | Value |
|---|---|---|---|
| $A_{mn}$ | dimensionless constant introduced in Eq. (26b) | | |
| $B_{mn}, B_{mn}^*$ | dimensionless constant introduced in Eq. (26a) | | |
| $Bi$ | Biot number, $hL/k_p$ | | |
| $c_{cf}$ | capacity of heat of fluid per unit mass | J kg$^{-1}$ K$^{-1}$ | |
| $c_p$ | capacity of heat of plate material per unit mass | J kg$^{-1}$ K$^{-1}$ | |
| $C_0, C_1 \ldots C_4$ | constants in normalized forms | | |
| $d_i$ | inside diameter of tubes to carryout the collector fluid | m | 0.014 |
| $D$ | dimensionless differential operator, see Eq. (24a) | | |
| $D_1, D_2$ | roots, defined in Eq. (24a) | | |
| $E_{mn}$ | constant term in normalized form, introduced in Eq. (44) | | |
| $F$ | dimensionless time, $\alpha t/L^2$ | | |
| $G, H$ | functions taken to execute the product method, see Eq. (21) | | |

*Corresponding author: E-mail: rahoolmech007@gmail.com.

| Symbol | Description | Units | Value |
|---|---|---|---|
| $h$ | convection coefficient | W m$^{-2}$ K$^{-1}$ | 300 |
| $i$ | complex number | | |
| $I, J$ | dimensionless constants | | |
| $L$ | length of the symmetric sector, see Fig. 1 | m | |
| $\dot{m}_{cf}$ | mass rate of flow of collector fluid | kg s$^{-1}$ | 0.2 |
| $k_p$ | thermal conductivity coefficient | W m$^{-1}$ K$^{-1}$ | |
| $S$ | solar irradiation | W m$^{-2}$ | |
| $S^*$ | solar flux in normalized form, defined in Eq. (3) | | |
| $t$ | time notation | s | |
| $t_p$ | plate thickness dimension | m | 0.0016 |
| $T_a$ | ambient temperature | °C | 20 |
| $T_{cf}$ | collector fluid temperature | °C | |
| $T_{fi}$ | inlet temperature of collector fluid | °C | 35 |
| $T_p$ | plate temperature | °C | |
| $U, V$ | functions employed for product method, see Eq. (16) | | |
| $U_l$ | overall loss coefficient | W m$^{-2}$ K$^{-2}$ | |
| $Ve$ | Vernotte number, $\sqrt{\alpha\tau/L^2}$ | | |
| $x, y$ | coordinates | m | |
| $X, Y$ | normalized coordinates ($x/L$, y/L) | | |
| $Z_0$ | plate thermal parameter, see Eq. (3) | | |

*Greek letters*

| Symbol | Description | Units | Value |
|---|---|---|---|
| $\alpha$ | thermal diffusivity of plate material, $k_p/\rho c_p$ | | |
| $\beta, \gamma$ | functions employed for product method, see Eq. (12) | | |
| $\delta, \phi$ | functions taken, see Eq. (7a) | | |
| $\theta$ | normalized plate temperature, $(T_p - T_a)/(T_{fi} - T_a)$ | | |
| $\theta_{cf}$ | dimensionless collector fluid temperature, $(T_{cf} - T_a)/(T_{fi} - T_a)$ | | |
| $\vartheta, f$ | variables introduced in Eq. (34) | | |
| $\lambda, \nu, \mu$ | Eigen constants in normalized forms | | |
| $\xi, \varphi$ | functions considered, see Eq. (32) | | |
| $\rho$ | absorber plate density | kg m$^{-3}$ | |
| $\sigma$ | transformed variable declared in Eq. (36) | | |
| $\varsigma$ | complex plane integral limit | | |
| $\tau$ | thermal relaxation time | s | |
| $\omega$ | temperature difference at the tube boundary of the absorber plate, $T_p - T_{cf}$ | °C | |
| $\Omega, \psi$ | functions adopted in product method, see Eq. (10) | | |

## 1. Introduction

The flat-plate solar collector is a thermal apparatus that absorbs solar radiation and converts it into useful heat energy for domestic household and space heating purposes and a variety of thermal applications. The collector consists of an absorber plate which is constructed by metal trips having high thermal conductivity and conducts the absorbed solar energy into the collector fluid. The performance of a solar collector is dependent significantly on the geometric specifications, and optical and thermophysical properties of the absorber plate. Due to the variable intensity of solar-radiation, unsteadiness always exists in conducting heat through the flat-plate collector. Therefore, it always demands to study the two-dimensional time-dependent behavior of flat plate collectors and is required to accurately predict the transient thermal response under different environmental conditions.

The steady-state and transient heat conduction analysis of solar collectors was done by many researchers. Duffie and Beckman[1] investigated the one-dimensional steady-state analysis of an absorber-plate for absorbing solar radiation by assuming the constant overall loss-coefficient, plate-thermal conductivity, and plate-thickness. Rahman et al.[2] carried out a more accurate analysis for conducting heat in the absorber plate based on the two-dimensional heat flow. Different absorber profile shapes were also considered to enhance the performance factor under steady conditions for the ease of manufacturing.[3–5] The steady-state analysis for estimating the collector performance is not at all suitable for the inherent variation of the solar flux absorbed with time by the plate. Wang et al.[6] evolved a new transient-method for the collector performance with varying solar radiance. The analysis of the collector fluid can also be lumped together with the absorber plate for better theoretical predictions of the performance of a solar-collector under unsteady weather conditions[7] and its dynamic behavior was studied by taking into account the change in solar incident flux received by the collector plate.[8] Emery and Rogers[9] developed a new transient method in analysis to predict the thermal performance characteristics of a collector from the solar irradiation measurement in variable conditions. Dhariwal and Mirdha[10] derived a closed-form expression for the thermal response behavior in transient conditions for the alteration of temperature along the direction of the collector-fluid flow. Two-dimensional unsteady heat conduction systems were solved by the finite difference method to estimate the temperature value at various discrete points.[11] Hamed et al.[12] modeled for a transient analysis to determine temperature responses of all components of a solar collector including the transparent cover and transfer fluid. A numerical study was conducted by Genc et al.[13] for the time-dependent behavior of a nanofluid based-flat-plate collector. A temperature-dependent thermal conductivity was assumed by Pasha et al.[14] for analyzing the unsteady heat transfer in an absorber-plate. Badescu et al.[15] evaluated a transient operation of solar air collectors in different radiative regimes taking into account a variable irradiation distribution. A simulation was performed to accurately estimate the long-term effect of a solar collector system using measured outdoor data with dynamic actions of direct and diffuse radiation.[16]

Apart from the absorber plate design parameters and variable solar radiation, the collector fluid also acts a significant role in the dynamic characteristic behavior of a solar-collector. The influence of the collector-fluid temperature on collector efficiency parameters was investigated through transient test methods by Amer et al.[17] Hilmer et al.[18] calculated the unsteady performance of a solar collector by changing the fluid mass-flow. Baughn and Young[19] determined the optimum fluid flow rate for the maximum collector efficiency for supplying hot water from a solar domestic-system. A 2D Fourier analysis of a solar-collector for varying the solar radiation with time was conducted by Onyegebabu and Morhenne[20] to determine the maximum exergetic efficiency and optimal flow rates. Collector fluid-flow as a function of time has a major effect on the energy collection in solar collector systems having large fluid volume and long connecting pipes.[21] Numerous researches were also conducted on the transient analysis of photovoltaic thermal (PVT) collectors which convert solar radiation into useable thermal energy and unused waste heat use to heat up the collector fluid.[22–25] Solar air heating and solar cooling devices are some of the applications of utilizing solar insolation into space heating and cooling. Hernandez et al.[26] numerically studied solar air heating collectors with porous matrixes to evaluate the thermo-energetic performance in transient state conditions. Using theoretical and experimental approaches, Hussein[27] investigated the unsteady condition of a flat-plate solar water-heater with thermosyphon effects. The performances of solar water heaters were investigated by Belessiotis and Mathioulakis[28] analytically to establish a heat conduction model. The influence of thermal capacitance on performances of a flat-plate collector for continuously variable climatic situations was executed by Klein et al.[29] Hamed et al.[30] determined effects of optimum operation parameters on the solar collector in transient behaviors using the exergy analysis. A transient FORTRAN simulation program of a nanofluid-based PVT system was developed by Kolahan et al.[31] Saroha et al.[32] conducted a theoretical analysis and experimenting of nanofluids-based solar photovoltaic/thermal hybrid collector. To enhance the heat transfer performance, Mahian et al.[33] used nanofluids in solar collectors. An improvement model was developed by Ibrahim et al.[34] to determine the instantaneous efficiency of a flat-plate solar thermal collector and this upgraded model is also suitable to employ when the system is off-operation.

Most of the aforementioned research works were studied using the classical law of heat conduction, which is infinite thermal wave propagation speed and heat flux directly proportional to the temperature gradient. It has been mentioned earlier that the accurate analysis of reliability and operational performance depends on the comprehension of the transient response of solar collectors in terms of thermal wave characteristics. Previously, the thermal analysis of absorber plates was done by utilizing the conventional heat conduction equation in parabolic nature under the assumption of moderate temperature gradients. However, the advent of advanced engineering technologies necessitates more innovative design and operational analysis. A sophisticated non-Fourier model of heat conduction based on a wave model is required for simulating a rapid transient heat transfer phenomena in two-dimensional flat-plate collectors. Han-Taw and Jae-Yuh[35] determined a non-Fourier heat transfer using an analytical approach. Many research works were employed different numerical schemes to calculate the hyperbolic heat transfer.[36–38] The exact solution of double lags in a wave mode of heat transfer was done by Zhang et al.[39] Kundu and Lee[40] demonstrated a comparative study between the classical and non-Fourier analyses for temperature distributions in a solar collector. An analytical closed-form expression for heat transfer satisfying the non-Fourier law in a finite medium was presented by Moosaie[41], which encompasses the effects of pulse surface heating and radiation. Many practical engineering problems were solved based on the finite thermal propagation.[42–44] Zima and Dziewa[45] modeled liquid flat-plate solar collector operating in transient states by considering a method with distributed parameters.

Recently Bellos and Tzivanidis[46] investigated a novel solar-driven cogeneration system with the combination of flat-plate collectors and evacuated tube collectors. Hashemi et al.[47] heated a residential building using two types of solar collectors, namely linear parabolic and flat solar collectors, and analyzed based on the technical and economical points of view. The thermal performance of flat-plate solar collectors with nanofluids was modeled by Sadeghzadeh et al.[48] using artificial intelligence techniques. Eltaweel et al.[49] investigated the enhancement of thermal performance of a flat-plate solar collector based on the twisted tube design heat exchanger filled with nanofluid.

The above literature review is based entirely on the one-dimensional classical and finite thermal propagation of heat in absorber-plates. Performances of a solar collector depend fundamentally on the overall temperature field produced which is a function of locations of the absorber plate at an instant time. Therefore, a slight variation in temperature distribution can change significantly the performance parameters in transferring heat. When heat transfer in a collector plate is not unidirectional due to its typical shape, and therefore, a 1-D approximation of the heat equation is not applicable to determine the exact temperature field in an absorber plate. This thermal condition is also occurred due to the variation of collector fluid temperature in the flow direction. Obviously, there are always possibilities for significant temperature variations along the y-direction in a large solar collector. In this situation, thermal responses in 2-D domain need to consider accomplishing an accurate model for performance analyses of a solar collector. Comprehension of two-dimensional heat transfer is to ensure the propagation speed inside the absorber plate. Motivated by the theory described above, an exact analytical solution in the 2-D Cartesian coordinates was illustrated in the present study for the transient response of two-dimensional temperature pattern in an absorber-plate. To develop a generalized model for the transient heat transfer, time lags based on a non-Fourier approach for the propagation of heat are adopted. This analysis is equally suitable for predicting the temperature response by the classical model to be chosen the zero value of thermal relaxation times. A finite propagation of solar energy in an absorber plate which is transferred to the collector fluid has been thoroughly examined as a function of Fourier number ($F$) and Vernotte number ($Ve$). An initial condition to run the solar collector system was assumed at a constant ambient temperature. The second type of unsteady temperature distribution has also been investigated when no further solar radiation falls on the plate or the solar collector is in stagnation condition. Under this situation, the absorber plate cools down from the steady-state condition to the dead condition and an analytical solution has been analyzed the transient phenomena under appropriate boundary conditions. In this case, a hybrid scheme comprises of the Laplace transform method (LTM) and product method to analytically solve the governing differential equation having a temperature as a function of coordinates. The present analytical solution has been compared with the previous studies to highlight the severe limitations of adopting the one-dimensional model for the transient analysis of the solar collector. A finite-difference numerical scheme has been adopted to further validate the present method and similar results indicate the development of an accurate analytical model in this study.

## 2. Mathematical Model

The transportation phenomenon of heat energy in solar collectors is a process involving absorption of incident solar energy, heat conduction to the collector fluid through the plate, and heat losses from absorber plate to surroundings. Thermal analysis is performed on the portion of the symmetric heat transfer section of an absorber plate having a single collector-fluid-carrying-tube as depicted in Fig. 1. For a small thickness taken, the temperature distribution in the plate is in two-directions and temperature variation along the plate thickness is negligibly small. The temperature of the collector-fluid rises in the flow direction. Some researchers[2, 20, 24, 35–38] had considered two-dimensional steady-state heat conduction in an absorber-plate to determine temperature fields. For an actual operating condition, there is always a variation of the rate of insolation day-to-day or even in a day with time. When absorber plates are kept under the solar radiation, as the heat propagation has a finite speed, it takes some time to achieve steady-state conditions. Hence, the temperature of an absorber-plate is not only a function of position but also dependent on time. A transient mathematical model is necessary at all times to determine the plate temperature under different transient conditions. Similarly, the solar collector stops its function when there is no movement of the collector fluid and no further solar radiation falls, signifying the transient state condition of the absorber plate. Most of the research works carried out using the classical law for heat conduction, which is heat flux as linearly dependent on the temperature gradient and thermal energy propagated at an infinite speed. In other words, a local change in temperature in the absorber plate resulting in instantaneous thermal change throughout the plate violates Einstein's theory of relativity. The classical heat transfer model has the limitation in analyzing practical engineering problems where very large thermal-gradient, low temperature, short-time or ultra-high heating speed are associated. Therefore, a thermal wave model for the heat transfer based on a finite propagation of speed was adopted by introducing a thermal relaxation time $\tau$. This model becomes advantageous as it also satisfies the classical model by selecting a very small or zero value of the relaxation time. Therefore, there may be no limitation for the propagation of heat in the plate associated with the present model.

*2.1. Assumptions*

The collector-plate heat transfer analysis was done by considering the following assumptions:

➢ No heat conduction takes place along the plate thickness $(t_p)$.

➢ Overall coefficient of heat loss $(U_l)$ and fluid specific heat $(c_p)$ are taken constants.

➢ Absorbed solar radiation $(S)$ has been taken as an instantaneous value.

➢ Flow-rate of collector fluid is invariant.

➢ Coefficient of conductivity $(k_p)$ of an absorber plate is unaltered owing to a low variation in temperature.

➢ The distance between two successive tubes is unaltered.

- Heat losses from plate-edges to surroundings are neglected due to a small thickness chosen.
- The initial state of the solar collector is assumed to be at ambient temperature ($T_a$).
- The difference between the plate-temperature adjacent to the fluid-carrying-tube and collector fluid temperature has been assumed a constant value ($\omega$).
- The width of the absorber plate is assumed to be half the length of the absorber plate.

Under these assumptions, the energy balances for a differential control volume with 2D heat flow in an absorber plate yields the following governing equation based on the thermal wave theory approach,

$$\frac{\partial}{\partial x}\left(k_p t_p \frac{\partial T_p}{\partial x}\right) + \frac{\partial}{\partial y}\left(k_p t_p \frac{\partial T_p}{\partial y}\right) + S - U_l(T_p - T_a) - \tau U_l \frac{\partial T_p}{\partial t} = \rho c_p \frac{\partial T_p}{\partial t} + \tau \rho c_p \frac{\partial^2 T_p}{\partial t^2} \tag{1}$$

where, the $T_p(x, y, t)$ is the local plate-temperature and $\rho$ is the mass density of the plate material. Constant surrounding temperature is denoted by $T_a$ and time $\tau$ represents a time lag for heat flow after an application of temperature gradient. Heat flow develops increasingly with the relaxation time of $\tau$, instead of the instantaneous propagation of heat after the absorber plate is heated up by the solar radiation. When the thermal relaxation time is a large value, the thermal disturbance spreads throughout the collector plate at finite velocities and it is a pattern of waves. The Fourier heat conduction can be determined from the above equation by limiting $\tau$ close to zero and the thermal heat diffusion occurs at the infinite speed.

The normalized form of the wave heat equation (Eq. 1) can be expressed as follows:

$$\frac{\partial^2 \theta}{\partial X^2} + \frac{\partial^2 \theta}{\partial Y^2} - Z_0^2 \theta + S^* = (1 + Z_0^2 Ve^2)\frac{\partial \theta}{\partial F} + Ve^2 \frac{\partial^2 \theta}{\partial F^2} \tag{2}$$

where

$$Z_0 = \sqrt{\frac{U_l L^2}{k_p t_p}}; \quad S^* = \frac{SL^2}{k_p t_p (T_{fi} - T_a)}; \quad X = \frac{x}{L}; \quad Y = \frac{y}{L}; \quad \theta = \frac{T_p - T_a}{T_{fi} - T_a}; \quad \alpha = \frac{k_p}{\rho c_p}; \quad F = \frac{\alpha t}{L^2}; \quad Ve = \sqrt{\frac{\alpha \tau}{L^2}} \tag{3}$$

In Eq. (3), the dimensionless temperature $\theta$ is defined by considering the collector fluid inlet temperature ($T_{fi}$) flowing inside the fluid-carrying tube adjacent to the boundary of the plate ($X = 1, Y = 0$). At the end of the tube ($X = 1, Y = 1$), the collector fluid temperature attains its highest value for the thermal energy absorbed continuously from absorber plates. Fourier number ($F$) and Vernotte number ($Ve$) represent the dimensionless time and heat flux propagation speed in the collector, respectively. When $Ve = 0$, the thermal propagation occurs instantaneously at the infinite speed and corresponds to the Fourier model. The non-Fourier model is valid for any other non-zero $Ve$ values. However, the non-Fourier aspect dominates at a high value of Ve. The thermal wave characteristics of a solar collector can be determined by utilizing initial and boundary conditions along with the governing equation. The heat flow in the flat-plate collector is always unsteady in nature during its starting and ending operations. At the beginning when solar flux is absorbed by a plate, the solar collector may reach to steady-state condition at a high value of operating time. On the other hand, after a period of time, when no further solar radiation falls on collector plates, absorber plate-temperature cools down from its steady-state to a dead condition. Hence, two different analytical solutions are presented to analyze the temperature distribution when the absorber is working under different operational conditions.

*2.2. Solution of transient model from starting of solar collector*

The determination of temperature response for two-dimensional heat wave requires two initial conditions and four boundary conditions. Considering that the initial state of the absorber plate is at ambient temperature, the normalized forms of the initial conditions can be represented as,

$$at\ F = 0: \begin{cases} \theta(X,Y,0)=0 \\ \partial\theta(X,Y,0)/\partial F=0 \end{cases} \quad (4)$$

The plate temperature becomes a maximum at its middle section between two consecutive tubes due to the symmetry which satisfies no transfer of solar energy across the section at $X = 0$. The boundary condition at $X = 1$ is considered for the transfer of solar energy from an absorber-plate to the collector-fluid with the local temperature ($T_{cf}$). Along the y-direction, the temperature gradient vanishes at $Y = 0$ and $Y = 1$ for $0 \leq X \leq 1$, satisfying the symmetric condition for the heat transfer propagation. Mathematically, boundary conditions in X- and Y-directions are given below in dimensionless forms:

$$\text{Along x} - \text{direction:} \quad \begin{aligned} X = 0: & \quad \frac{\partial\theta(0,Y,F)}{\partial X} = 0 \\ X = 1: & \quad \frac{\partial\theta(1,Y,F)}{\partial X} = -Bi(\theta - \theta_{cf}) \end{aligned} \quad (5a)$$

$$\text{Along y} - \text{direction:} \quad \begin{aligned} Y = 0: & \quad \frac{\partial\theta(X,0,F)}{\partial Y} = 0 \\ Y = 1: & \quad \frac{\partial\theta(X,1,F)}{\partial Y} = 0 \end{aligned} \quad (5b)$$

Here, a normalized fluid temperature $\theta_{cf}$ is related to the dimensional fluid temperature $T_{cf}$ and collector fluid inlet temperature ($T_{fi}$) as described in Eq. (6c). When the collector fluid receives the useful heat from the plate, the heat transfer encounters a thermal resistance due to the bond and tube. The Biot number based on the collector fluid is defined as $Bi = hL/k_p$. The fluid temperature expressed as a function of y-coordinate can be obtained by solving a governing energy equation of working fluid inside the tube. The collector fluid-inlet temperature was slightly higher than the ambient temperature is taken and it was a design constant. Hence, we have

$$-\dot{m}_{cf}c_{cf}\frac{dT_{cf}}{dy} + \pi d_i h\big(T_p(L,y,t) - T_{cf}\big) = 0 \quad (6a)$$

$$T_{cf} = T_{fi} \quad at\ y = 0 \quad (6b)$$

In Eq. (6a), the specifications of the collector tube are given by its inner diameter $d_i$ and the mass flow rate $\dot{m}_{cf}$.. Eq. (6a) can be recast into their dimensionless form and solved to obtain the dimensionless fluid temperature as

$$\theta_{cf} = \frac{T_{cf} - T_a}{T_{fi} - T_a} = 1 + \frac{\pi d_i k_p Bi\omega}{\dot{m}_{cf} c_{cf}(T_{fi} - T_a)} Y \quad (6c)$$

Here, it can be observed that $\theta_{cf}$ is a linear relationship with $Y$. The temperature pattern in an absorber plate is determined by a solution of Eq. (2) using initial and boundary conditions (Eqs. 4 and 5). Another point to note is that only one initial condition is required for determining temperature using Fourier heat conduction. To apply a product method[50] in the case of the 2-D non-Fourier analysis in Eq. (2), the following set of variables can be written.

$$\theta(X,Y,F) = \phi(X,Y,F) + \delta(X,Y) \quad (7a)$$

$$\frac{\partial^2 \phi}{\partial X^2} + \frac{\partial^2 \phi}{\partial Y^2} - Z_0^2 \phi = (1 + Z_0^2 Ve^2)\frac{\partial \phi}{\partial F} + Ve^2 \frac{\partial^2 \phi}{\partial F^2} \quad (7b)$$

$$\frac{\partial^2 \delta}{\partial X^2} + \frac{\partial^2 \delta}{\partial Y^2} - Z_0^2 \delta + S^* = 0 \quad (7c)$$

The boundary conditions for the solution of Eqs. (7b) and (7c) can be obtained by combining Eqs. (7a), (4), and (5) as follows:

$$\phi(X,Y,0) = -\delta(X,Y); \quad \frac{\partial \phi(X,Y,0)}{\partial F} = 0 \ (0 \leq X,Y \leq 1) \quad (8)$$

$$\frac{\partial \phi(0,Y,F)}{\partial X} = 0; \quad \frac{\partial \phi(1,Y,F)}{\partial X} + Bi\phi(1,Y,F) = 0; \quad \frac{\partial \phi(X,0,F)}{\partial Y} = \frac{\partial \phi(X,1,F)}{\partial Y} = 0 \quad (9a)$$

$$\frac{\partial \delta(0,Y)}{\partial X} = 0; \quad \frac{\partial \delta(1,Y)}{\partial X} = -Bi(\delta(1,Y) - \theta_{cf}); \quad \frac{\partial \delta(X,0)}{\partial Y} = \frac{\partial \delta(X,1)}{\partial Y} = 0 \quad (9b)$$

Here, it is to note that the determination of the closed-form solution for Eq. (7c) with the boundary conditions given in Eq. (9b) is necessary to find a set of variables that satisfies the non-homogeneity of the boundary condition in the x-direction along with the symmetric condition in the y-direction. Hence, the following set of variables are implemented and substituted in Eq. (7c) as,

$$\delta(X,Y) = \Omega(X) + \psi(X,Y) + e^{-Z_0 X} + e^{-Z_0 Y} \quad (10)$$

$$\frac{d^2\Omega}{dX^2} - Z_0^2 \Omega + S^* = 0 \quad (11a)$$

$$\frac{\partial^2 \psi}{\partial X^2} + \frac{\partial^2 \psi}{\partial Y^2} - Z_0^2 \psi = 0 \quad (11b)$$

Separation of variables can again be applied in Eq. (11b) to obtain two ODEs as,

$$\psi(X,Y) = \beta(X)\gamma(Y) \quad (12a)$$

$$\frac{d^2\beta}{dX^2} - \frac{Z_0^2 \beta}{2} = 0; \quad \frac{d^2\gamma}{dY^2} - \frac{Z_0^2 \gamma}{2} = 0 \quad (12b)$$

The solution of Eqs. (11a) and (11b) yields,

$$\Omega(X) = S^*/Z_0^2 + \sinh(Z_0 X) - C_0 \cosh(Z_0 X) \quad (13a)$$

$$\psi(X,Y) = [C_1 \cosh(Z_0 X/\sqrt{2}) + C_3 \sinh(Z_0 X/\sqrt{2})][C_2 \cosh(Z_0 Y/\sqrt{2}) + C_4 \sinh(Z_0 Y/\sqrt{2})] \quad (13b)$$

where

$$C_0 = \left[\frac{Bi\{S^*/Z_0^2 + \sinh(Z_0)\} + Z_0 \sinh(Z_0)}{Bi\cosh(Z_0) + Z_0 \sinh(Z_0)}\right]$$

$$C_1 = \frac{(\theta_{cf} - e^{-Z_0})}{\cosh(Z_0/\sqrt{2}) + Z_0/(Bi\sqrt{2})\sinh(Z_0/\sqrt{2})}$$

$$C_2 = \frac{\sqrt{2}(e^{-Z_0} - \cosh(Z_0/\sqrt{2})}{\sinh(Z_0/\sqrt{2})}$$

(14)

Finally, the solution of Eq. (7c) can be obtained from Eqs. (10), (13a), and (13b) as follows:

$$\delta(X,Y) = S^*/Z_0^2 + \sinh(Z_0 X) - C_0 \cosh(Z_0 X) + C_1 \cosh\left(\frac{Z_0 X}{\sqrt{2}}\right)$$

$$\left[C_2 \cosh\left(\frac{Z_0 Y}{\sqrt{2}}\right) + \sqrt{2}\sinh(Z_0 Y/\sqrt{2})\right] + e^{-Z_0 X} + e^{-Z_0 Y}$$

(15)

To solve Eq. (7b) for the variable $\phi(X,Y,F)$, the following functions can be taken explicitly:

$$\phi(X,Y,F) = U(X,Y)V(F) \tag{16}$$

The governing Eq. (7b) can then be written using Eq. (16) as

$$\frac{\partial^2 U}{\partial X^2} + \frac{\partial^2 U}{\partial Y^2} + \lambda^2 U = 0 \tag{17}$$

$$Ve^2 \frac{d^2 V}{dF^2} + (1 + Z_0^2 Ve^2)\frac{dV}{dF} + (Z_0^2 + \lambda^2)V = 0 \tag{18}$$

The above differential equations are obtained by assuming an Eigen constant $\lambda$. Initial and boundary conditions expressed in Eqs. (8) and (9a) can be transformed as,

$$U(X,Y)V(0) = -\delta(X,Y); \quad \frac{dV(0)}{dF} = 0 \tag{19}$$

$$\frac{\partial U(0,Y)}{\partial X} = 0; \quad \frac{\partial U(1,Y)}{\partial X} + BiU(1,Y) = 0; \quad \frac{\partial U(X,0)}{\partial Y} = \frac{\partial U(X,1)}{\partial Y} = 0 \tag{20}$$

Eq. (17) can be further modified by considering $U(X,Y)$ as a product of two variables $G(X)$ and $H(Y)$. From Eqs. (17) and (20), we get the following mathematical relations.

$$\frac{d^2 G}{dX^2} + v^2 G = 0 \; ; \; \frac{d^2 H}{dY^2} + (\lambda^2 - v^2)H = 0 \tag{21}$$

$$\frac{dG(0)}{dX} = 0; \quad \frac{dG(1)}{dX} + BiG(1) = 0; \quad \frac{dH(0)}{dY} = \frac{dH(1)}{dY} = 0 \tag{22}$$

The solution for Eq. (17) can be obtained using Eq. (19) based on the boundary conditions given in Eq. (22) as

$$U_{m,n}(X,Y) = G_m(X)H_n(Y) = \sum_{m=0}^{\infty}\sum_{n=0}^{\infty} E_{mn}\cos(v_m X)\cos(\mu_n Y) \tag{23a}$$

$$v_m = \tan^{-1}\frac{Bi}{v_m} + m\pi; \quad \mu_n = n\pi \; ; \; \lambda_{mn}^2 = v_m^2 + \mu_n^2 \tag{23b}$$

Eq. (18) can be rewritten using a differential operator $D$, which can be determined as

$$[Ve^2 D^2 + (1 + Z_0^2 Ve^2)D + (Z_0^2 + \lambda_{mn}^2)]V = 0$$

$$D = \{D_1, D_2\} = \frac{-(1 + Z_0^2 Ve^2) \pm \sqrt{(1 + Z_0^2 Ve^2)^2 - 4Ve^2(Z_0^2 + \lambda_{mn}^2)}}{2Ve^2} \tag{24a}$$

The values of $D_1$ and $D_2$ in Eq. (23a) can be expressed by the following constants.

$$I = \frac{-(1 + Z_0^2 Ve^2)}{2Ve^2}; \quad J = \frac{\sqrt{(1 + Z_0^2 Ve^2)^2 - 4Ve^2(Z_0^2 + \lambda_{mn}^2)}}{2Ve^2} \tag{24b}$$

From Eq. (24), $D$ can be real, zero or imaginary depending upon the situation. Eqs. (23a) and (24) can be combined together in Eq. (16) to obtain the solution for $\phi(X, Y, F)$ as

$$\phi(X, Y, F) = \sum_{m=0}^{\infty} \sum_{n=0}^{\infty} (B_{mn} e^{D_1 F} + B_{mn}^* e^{D_2 F}) \cos(\nu_m X) \cos(\mu_n Y) \tag{25}$$

The initial conditions in Eq. (19) can be used to find the constants $B_{mn}$ and $B_{mn}^*$ in Eq. (25) as

$$\phi(X, Y, 0) = \sum_{m=0}^{\infty} \sum_{n=0}^{\infty} (B_{mn} + B_{mn}^*) \cos(\nu_m X) \cos(\mu_n Y) = -\delta(X, Y) \tag{25a}$$

$$\frac{\partial \phi(X, Y, 0)}{\partial F} = \sum_{m=1}^{\infty} \sum_{n=1}^{\infty} (B_{mn} D_1 + B_{mn}^* D_2) \cos(\nu_m X) \cos(\mu_n Y) = 0 \tag{25b}$$

It is clear that Eq. (25a) is a double Fourier series of $\delta(X, Y)$ and can be solved by using the generalized Euler formula. Then, the values for the constants $B_{mn}$ and $B_{mn}^*$ can be determined as

$$B_{mn}^* = -B_{mn} \frac{D_1}{D_2}; \quad B_{mn} = -4 \frac{D_2 A_{mn}}{D_2 - D_1} \tag{26a}$$

$$A_{mn} = \left[ C_1 C_3 \left( \frac{C_2(Z_0/\sqrt{2})\sinh(Z_0/\sqrt{2})\cos(\mu_n) + Z_0\cosh(Z_0/\sqrt{2})\cos(\mu_n) - Z_0}{(Z_0^2/2) + \mu_n^2} \right) + C_4 \right] \tag{26b}$$

where

$$C_3 = \frac{(Z_0/\sqrt{2})\sinh(Z_0/\sqrt{2})\cos(\nu_m) + \nu_m \cosh(Z_0/\sqrt{2})\sin(\nu_m)}{(Z_0^2/2) + \nu_m^2}; \quad C_4 = \frac{Z_0 e^{-Z_0}(1 - \cos(\mu_n))\sin(\nu_m)}{(Z_0^2 + \mu_n^2)\nu_m} \tag{27}$$

Eqs. (25) was integrated in combination with the prescribed set of variables in Eqs. (15) and (25), and the temperature distribution from the non-Fourier model with the help of the separation of variables can be obtained as

For $4Ve^2(Z_0^2 + \lambda_{mn}^2) < (1 + Z_0^2 Ve^2)^2$:

$$\theta(X, Y, F) = S^*/Z_0^2 + \sinh(Z_0 X) - [C_0] \cosh(Z_0 X)$$
$$+ C_1 \cosh(Z_0 X/\sqrt{2}) [C_2 \cosh(Z_0 Y/\sqrt{2}) + \sqrt{2} \sinh(Z_0 Y/\sqrt{2})] + e^{-Z_0 X} + e^{-Z_0 Y}$$
$$- 4 \sum_{m=0}^{\infty} \sum_{n=0}^{\infty} A_{mn} \exp(IF) \left( \cosh(JF) - \frac{I}{J} \sinh(JF) \right) \cos(\nu_m X) \cos(\mu_n Y) \tag{28a}$$

For $4Ve^2(Z_0^2 + \lambda_{mn}^2) = (1 + Z_0^2 Ve^2)^2$:

$$\theta(X,Y,F) = S^*/Z_0^2 + \sinh(Z_0 X) - [C_0]\cosh(Z_0 X)$$
$$+ C_1\cosh(Z_0 X/\sqrt{2})\left[C_2\cosh(Z_0 Y/\sqrt{2}) + \sqrt{2}\sinh(Z_0 Y/\sqrt{2})\right] + e^{-Z_0 X} + e^{-Z_0 Y}$$
$$- 4\sum_{m=0}^{\infty}\sum_{n=0}^{\infty} A_{mn}\exp(IF)(1 - IF)\cos(v_m X)\cos(\mu_n Y) \quad (28b)$$

For $4Ve^2(Z_0^2 + \lambda_{mn}^2) > (1 + Z_0^2 Ve^2)^2$:

$$\theta(X,Y,F) = S^*/Z_0^2 + \sinh(Z_0 X) - [C_0]\cosh(Z_0 X)$$
$$+ C_1\cosh(Z_0 X/\sqrt{2})\left[C_2\cosh(Z_0 Y/\sqrt{2}) + \sqrt{2}\sinh(Z_0 Y/\sqrt{2})\right] + e^{-Z_0 X} + e^{-Z_0 Y}$$
$$- 4\sum_{m=0}^{\infty}\sum_{n=0}^{\infty} A_{mn}\exp(IF)\left(\cos(JF) - \frac{I}{J}\sin(JF)\right)\cos(v_m X)\cos(\mu_n Y) \quad (28c)$$

A normalized form of energy equation formulated by applying the classical heat conduction law is written by putting $Ve = 0$ in Eq. (2).

$$\frac{\partial^2 \theta}{\partial X^2} + \frac{\partial^2 \theta}{\partial Y^2} - Z_0^2 \theta + S^* = \frac{\partial \theta}{\partial F} \quad (29)$$

Eq. (29) is solved by adopting the same steps of the above analysis for the non-Fourier analysis and considering one initial and four boundary conditions. Therefore, the 2-D temperature profile based on the classical heat conduction law can be represented as

$$\theta(X,Y,F) = S^*/Z_0^2 + \sinh(Z_0 X) - [C_0]\cosh(Z_0 X)$$
$$+ C_1\cosh(Z_0 X/\sqrt{2})\left[C_2\cosh(Z_0 Y/\sqrt{2}) + \sqrt{2}\sinh(Z_0 Y/\sqrt{2})\right] + e^{-Z_0 X} + e^{-Z_0 Y}$$
$$- 4\sum_{m=0}^{\infty}\sum_{n=0}^{\infty} A_{mn}\exp[-(Z_0^2 + \lambda_{mn}^2)F]\cos(v_m X)\cos(\mu_n Y) \quad (30)$$

*2.3. Solution of transient model during stagnation of solar collector*

In the previous section, it has been discussed about the transient behavior of the solar collector when it starts operating until it reaches the steady-state condition. However, in practical applications, its smooth operation may get interrupted when the absorber plate receives no more solar radiation from the sun due to overcast weather conditions. Then the absorber plate cools down from its steady-state temperature to the ambient temperature. When the collector fluid reaches the maximum temperature early in the day or its flow gets interrupted, the solar heating system gets overheated when further solar radiation is absorbed by it resulting in a condition commonly called "stagnation". Overheating of the solar collector causes significant damage to its components over time if proper preventive measures are not undertaken. It is necessary to determine the plate temperature of a solar-collector during the time taken by it to cool down below the operating temperature, so that an appropriate protection method can be designed to accomplish a successful control of the cooling process without damaging its components. The transient thermal response of a collector plate during the entire process can be analyzed by adopting the following initial and boundary conditions.

$$\frac{\partial^2 \theta}{\partial X^2} + \frac{\partial^2 \theta}{\partial Y^2} - Z_0^2 \theta = (1 + Z_0^2 Ve^2)^2 \frac{\partial \theta}{\partial F} + Ve^2 \frac{\partial^2 \theta}{\partial F^2} \tag{31a}$$

$$I.C.: \quad \theta(X,Y,0) = \theta_s(X,Y); \quad \frac{\partial \theta(X,Y,0)}{\partial F} = 0 \tag{31b}$$

$$B.C.: \quad \frac{\partial \theta(0,Y)}{\partial X} = \frac{\partial \theta(1,Y)}{\partial X} = 0; \frac{\partial \theta(X,0)}{\partial Y} = \frac{\partial \theta(X,1)}{\partial Y} = 0 \tag{31c}$$

Eq. (31a) can be easily solved by Laplace Transform Method (LTM)[51] considering an initial condition of the absorber plate is at steady-state and the boundary conditions are of Neumann type. First, the variable $\theta(X,Y,F)$ is separated by using the following set of variables.

$$\theta(X,Y,F) = \xi(X,F)\varphi(Y,F) \tag{32}$$

Eq. (31a) can be modified by taking $Ve = 0$ for Fourier analysis, and split up by the above equation as follows.

$$\frac{\partial^2 \xi}{\partial X^2} - \frac{Z_0^2}{2}\xi - \frac{\partial \xi}{\partial F} = 0 \tag{33a}$$

$$\frac{\partial^2 \varphi}{\partial Y^2} - \frac{Z_0^2}{2}\varphi - \frac{\partial \varphi}{\partial F} = 0 \tag{33b}$$

Let us assume $\overline{f}(X) = \xi(X,0)$ and $\overline{\xi}(X,s) = \vartheta(X)$. Apply LTM to Eq. (33a) and the boundary conditions:

$$\frac{\partial^2 \overline{\xi}(X,s)}{\partial X^2} - \left(\frac{Z_0^2}{2} + s\right)\overline{\xi}(X,s) = -\overline{f}(X) \tag{34}$$

$$\overline{\xi}(0,s) = \vartheta'(0) = 0 \tag{35a}$$

$$\overline{\xi}(1,s) = \vartheta'(1) = 0 \tag{35b}$$

To find $\vartheta(X)$ with the transformed variable $\sigma$, we apply LTM to Eq. (34) and then inverse theorem, we get the transformed function as,

$$\vartheta(X) = L^{-1}\left[\frac{\sigma \vartheta(0)}{\sigma^2 - (Z_0^2/2 + s)} - \frac{\overline{f}(\sigma)}{\sigma^2 - (Z_0^2/2 + s)}\right] \tag{36}$$

Eq. (36) can be simplified by implementing the transformed boundary conditions Eq. (35) after taking the inverse as

$$\vartheta(X) = \int_{u=0}^{X} \overline{f}(u) \frac{\cosh\left(\sqrt{(Z_0^2/2 + s)}u\right)\cosh\left(\sqrt{(Z_0^2/2 + s)}(1-X)\right)}{\sqrt{(Z_0^2/2 + s)}\sinh\left(\sqrt{(Z_0^2/2 + s)}\right)} du$$

$$+ \int_{u=X}^{1} \overline{f}(u) \frac{\cosh\left(\sqrt{(Z_0^2/2 + s)}X\right)\cosh\left(\sqrt{(Z_0^2/2 + s)}(1-u)\right)}{\sqrt{(Z_0^2/2 + s)}\sinh\left(\sqrt{(Z_0^2/2 + s)}\right)} du \tag{37}$$

In the complex plane, the inverse theorem (residue form) is applied to Eq. (37) to get the following expressions:

$$\xi(X,F) = \frac{1}{2\pi i}\lim_{L\to\infty}\int_{\vartheta-iL}^{\vartheta+iL} e^{sF}\overline{\xi}(X,s) \tag{38}$$

$$\sum Re\,s = \int_{x_0-i\infty}^{x_0+i\infty} e^{sF}\left[\int_{u=0}^{X} \overline{f}(u)\frac{\cosh\left(\sqrt{(Z_0^2/2+s)}u\right)\cosh\left(\sqrt{(Z_0^2/2+s)}(1-X)\right)}{\sqrt{(Z_0^2/2+s)}\sinh\left(\sqrt{(Z_0^2/2+s)}\right)}du\right.$$
$$\left.+\int_{u=X}^{1} \overline{f}(u)\frac{\cosh\left(\sqrt{(Z_0^2/2+s)}X\right)\cosh\left(\sqrt{(Z_0^2/2+s)}(1-u)\right)}{\sqrt{(Z_0^2/2+s)}\sinh\left(\sqrt{(Z_0^2/2+s)}\right)}du\right]ds \quad (39)$$

The residue can be obtained from Eq. (39) and subsequently be solved by taking the limits when tends to a residue value

$$s = -(m^2\pi^2 + Z_0^2/2) \quad (40)$$

$Re\,s[-(m^2\pi^2+Z_0^2/2)]$

$$=\lim_{s\to-(m^2\pi^2+Z_0^2/2)}[s+m^2\pi^2+Z_0^2/2]\left[\int_{u=0}^{X} \overline{f}(u)\frac{\cosh\left(\sqrt{(Z_0^2/2+s)}u\right)\cosh\left(\sqrt{(Z_0^2/2+s)}(1-X)\right)}{\sqrt{(Z_0^2/2+s)}\sinh\left(\sqrt{(Z_0^2/2+s)}\right)}du\right.$$
$$\left.+\int_{u=X}^{1} \overline{f}(u)\frac{\cosh\left(\sqrt{(Z_0^2/2+s)}X\right)\cosh\left(\sqrt{(Z_0^2/2+s)}(1-u)\right)}{\sqrt{(Z_0^2/2+s)}\sinh\left(\sqrt{(Z_0^2/2+s)}\right)}du\right]e^{sF} \quad (41)$$

$$\xi(X,F) = 2\sum_{m=0}^{\infty} e^{-(m^2\pi^2+Z_0^2/2)F}\int_0^1 [\overline{f}(u)\cos(m\pi u)du]\cos(m\pi X) \quad (42)$$

Following the same procedure, we can solve Eq. (33b) to get $\varphi(Y,F)$. Then, from Eq. (32) we get,

$$\theta(X,Y,F) = 4\sum_{m=0}^{\infty}\sum_{n=0}^{\infty} e^{-(m^2\pi^2+n^2\pi^2+Z_0^2)F}\int_0^1[\overline{f}(u)\cos(m\pi u)\cos(n\pi u)du]\cos(m\pi X)\cos(n\pi Y) \quad (43)$$

From the initial conditions given in Eq. (31b), we can solve Eq. (43) by integration and the temperature distribution is expressed as

$$\theta(X,Y,F) = 4\sum_{m=0}^{\infty}\sum_{n=0}^{\infty} E_{mn}e^{-(m^2\pi^2+n^2\pi^2+Z_0^2)F}\cos(m\pi X)\cos(n\pi Y) \quad (44a)$$

$$E_{mn} = \left[\frac{C_1(Z_0/\sqrt{2})\sinh(Z_0/\sqrt{2})\cos(m\pi)}{(Z_0^2/2)+m^2\pi^2}\right]$$
$$\times\left[\frac{C_2(Z_0/\sqrt{2})\sinh(Z_0/\sqrt{2})\cos(n\pi)+Z_0\cosh(Z_0/\sqrt{2})\cos(n\pi)-Z_0}{(Z_0^2/2)+n^2\pi^2}\right] \quad (44b)$$

Based on the same mathematical formulations, the temperature distribution for the non-Fourier condition obtained by solving Eq. (31) as

For $(m^2\pi^2 + n^2\pi^2 + Z_0^2) < (1+Z_0^2Ve^2)^2/4Ve^2$:

$$\theta(X,Y,F) = 4\sum_{m=0}^{\infty}\sum_{n=0}^{\infty} E_{mn}\exp(IF)\left(\cosh(JF)-\frac{I}{J}\sinh(JF)\right)\cos(m\pi X)\cos(n\pi Y) \quad (45a)$$

For $(m^2\pi^2 + n^2\pi^2 + Z_0^2) = (1 + Z_0^2 Ve^2)^2/4Ve^2$:

$$\theta(X,Y,F) = 4 \sum_{m=0}^{\infty} \sum_{n=0}^{\infty} E_{mn} \exp(IF)(1 - IF) \cos(m\pi X) \cos(n\pi Y) \qquad (45b)$$

For $(m^2\pi^2 + n^2\pi^2 + Z_0^2) > (1 + Z_0^2 Ve^2)^2/4Ve^2$:

$$\theta(X,Y,F) = 4 \sum_{m=0}^{\infty} \sum_{n=0}^{\infty} E_{mn} \exp(IF) B_1 \left(\cos(JF) - \frac{I}{J}\sin(JF)\right) \cos(m\pi X) \cos(n\pi Y) \qquad (45c)$$

where

$$J = \frac{\sqrt{\left|\left(1 + Z_0^2 Ve^2\right)^2 - 4Ve^2\left(m^2\pi^2 + n^2\pi^2 + Z_0^2\right)\right|}}{2Ve^2} \qquad (46)$$

The absorber plate temperature under the 2-D heat flow expressed in Eq. (30) and Eq. (45) are an infinite convergence series. The exact analytical solution will have no error if we consider the infinite number of terms. In the present paper, we have considered around 40 terms to determine the temperature response without any error up to 4 places of decimal.

## 3. Results and possible discussion

An accurate determination of heat transmission in a solar collector is complex due to the finite speed of propagation of heat in the absorber plate along with the rapidly changing weather conditions. The seasonable effect changes the rate of solar irradiation. To determine the appropriate performance of a solar collector, researchers have implemented transient physical modeling for its temperature evolution. A closed-form solution for a normalized 1-D temperature pattern in both the classical and wave approaches was reported by Kundu and Lee[40] with a theoretical work for an initial ambient temperature. However, the major shortcoming of this model is that based on the assumption of one-dimensional heat conduction and no temperature variation considered along the direction of collector fluid flow. Based on the available literature, two-dimensional analysis in the absorber-plate has not yet been performed using the heat propagation in wavy character. The primary aim of this work is to investigate a deviation of the present analysis from the published 1-D work and the present two-dimensional approach is a more accurate evaluation of the temperature distribution under different initial and boundary conditions.

### 3.1. Validation of the present work

According to the literature survey, no practical work exists on the 2-D temperature field of an absorber plate under solar radiation. Hence, the validation of the present thermal response can be made with the 1-D non-Fourier temperature determined by Kundu and Lee.[40] In Fig. 2a, the comparison of temperature pattern as a functional relationship with the x-coordinate at various Fourier numbers between the present analysis and the published work has been highlighted. These results were obtained by satisfying a convected condition at the boundary normal to y-direction and the collector fluid temperature throughout the tube is assumed constant at $T_{fi}$. For the comparative study, design properties and solar radiation value are set as per the published work for $Z_0 = 0.5, Bi = 0.5, Ve = 0.2,$ and $S^* = 1.0$. At any instant, the temperature pattern for both classical and thermal wave laws is a coordinate-dependent. Hence, an average temperature pattern ($\theta_{av} = \int_{Y=0}^{1} \theta dY$) in the y-direction has been plotted as a sole function of x-coordinate. A slight deviation in the temperature curve has been observed for the 2-D based non-Fourier model due to the variation in temperature along the y-direction. However, this disparity minimizes at a higher Fourier number where the predicted temperature distributions of the present model are in agreement with the previous 1-D model. Fig. 2b depicts a comparative temperature as a relation of normalized time term ($F$) between the present work and the published work for both $S^* = 1.0$ and $S^* = 2.0$ at a section ($X = 0.5$) with $Ve = 0.5$. As depicted in the figure, the present work produces a minor variation in the temperature response at the initial time owing to the heat flow that occurs in two directions and the involvement of the hyperbolic nature for energy propagation. However, the scope of this deviation between 1-D and 2-D thermal wave models is completely dependent on the Fourier number and thermal relaxation time. The zero deviation is achievable when the absorber plate gradually attains a thermal equilibrium with a further increase in Fourier value. For $S^* = 1.0$, the maximum temperature is $\theta_{av} = 2.08$, and for $S^* = 2.0$, it is $\theta_{av} = 3.71$. From the figure, it is clear that the published work reported lower temperatures than the present study. Therefore, the mathematical modeling of heat transfer in solar collectors is required to modify in considering the 2-D non-Fourier model for the correct temperature response.

The analytical solution for the Fourier and non-Fourier 2-D model can be further validated by solving Eq. (2) numerically with the finite difference method. Using Taylor's theorem, the governing partial differential equation has been uniformly discretized with the first-order forward difference in time and the central difference for the space derivative terms of second-order accuracy. As the boundary conditions are of either Neumann or Robin type, the temperature derivative at the boundary is expressed by introducing ghost cells outside the physical domain. An explicit method was adopted for solving the difference equations at different time step and at different equidistant grid points in the absorber plate. The zero initial condition is used to find the temperature values for the first time step. The discretization error is kept as minimum as possible and the stability condition is maintained at every iteration. The detailed numerical method can be found in previous literature.[11] Figs. 3a and 3b demonstrate a thermal response based on the 2-D Fourier model and the non-Fourier model, respectively in solar collectors along the x-direction for the variation of Fourier number from 0.01 to 10. Rate of solar incident radiation on the absorber is taken as $S^* = 1$.

Temperature distributions were plotted for $T_{fi} = 35°C$ at a location $(Y = 0.5)$ with $Ve = 1.5$. Associated design variables are set as $Z_0 = 0.5$ and $Bi = 0.5$, respectively. The initial condition of the plate is at ambient temperature $(T_a = 20°C)$. As $F$ increases, the plate temperature is to rise due to more solar energy absorbed and finally, energy is transferred to the collector fluid. Temperatures of the plate at the midsection between two consecutive tubes are always greater than the temperature of the plate near the tube region. The thermal energy of the collector fluid increases along the y-direction as the useful heat gained occurs from the absorber plate. It can be observed that the difference between the temperature at the line of symmetric and the temperature at the tube region is higher at the inlet end $(Y = 0)$ than at the outlet end $(Y = 1)$. Furthermore, the rate of temperature does not increase uniformly with increasing $F$. Under this situation, the non-Fourier temperature increases significantly at low values of $F$, but afterward the thermal response follows a declining nature until it reaches the maximum equilibrium temperature. Fig. 3a depicts the thermal response of the Fourier law. In the absence of $Ve$, the magnitude of temperature distribution for a Fourier condition is much higher at low $F$ values compared to the non-Fourier analysis in Fig. 3b. With further increase in $F$, the temperature profile in the plate does not change with the $F$. After $F = 10.0$, both the classical and thermal wave temperatures reach an equilibrium. The implication for the departure of the temperature response involves the thermal relaxation time in the governing equation as derived in Eq. (1). For the lagging time for the propagation of heat $(Ve = 1.5)$, the temperature response for a wavy nature of heat transfer is lesser than the classic mode where the lagging phenomenon is absent. This event truly shows the justification of the thermal wave in nature of heat transfer in the analysis. Above a certain $F$ value, it is observed that the plate-temperature decreases in the heat flow direction and it is higher than the fluid temperature. This is obviously owing to the heat flow from the hot plate to the cold collector fluid. Furthermore, patterns of the temperature in the physical domain satisfy clearly the boundary conditions associated with the present analysis in the x-direction. Figs. 3c and 3d show the temperature variation as a function of $Y$ at $X = 0.5$ for both classical and thermal wave models, respectively. The energy transport to the collector fluid from the plate occurs having a negative temperature gradient. At any location of the plate, the temperature pattern is governed by the local temperature of the fluid whereas an increasing temperature has been observed with the increase in time. For the validation purpose, the numerically obtained temperature distribution for both the classical and thermal wave models has been depicted in Fig. 3. In all these figures, the numerical results are comparable magnitude with the temperature pattern determined by the analytical methods. It is to note that if the Vernotte number is taken as zero directly in the non-Fourier temperature formulated in Eq. (28), then the thermal wave model converts into the classical model. But this approach may lead to infinity in the mathematical calculations. On the other hand, the magnitude of $\theta$ is always higher for an increasing $Z_0$ but the temperature pattern doesn't change as depicted in Fig. 3.

*3.2. Temperature response of the solar collector under starting condition*

Since the present analysis investigates the temperature response in a flat-plate collector having two-dimensional, isotherm lines can give an appropriate idea. Fig. 4 shows the isotherm lines of an absorber plate for different Fourier numbers based on the thermal wave model ($Ve = 0.5$). Fig. 4a depicts the 2-D temperature plot just after the starting of the solar collector at $F = 0.01$. The contour plot clearly shows the direction of heat flow in the examined domain of the absorber plate. The temperature gradually reduces towards the collector fluid-carrying tube in both x- and y- directions and the isotherms are a unique pattern with respect to the collector tube. Owing to have a finite speed of thermal propagation, an irregularly shaped temperature variation has been observed from the graphical plot. The orientation of the 2-D temperature graph satisfies the Neumann type of boundary conditions on three sides of the collector plate. With further increase in Fourier value, a considerable deviation in the thermal curve orientation has been noticed compared to Fig. 4a. The 2-D temperature contour for $F = 0.5$ is represented in Fig. 4b, and an increment in temperature pattern is observed in the y-direction. In the case of $F = 1.0$ and $F = 2.5$, referring to Figs. 4c and 4d, a similar trend is found with more number of contour curves in the 2-D examined domain. It highlights that at a high Fourier number, constant temperature lines are gradually becoming normal to the collector tube, which implies that a considerable part of heat flux is parallel to the tube. The rate of temperature variation decreases with an increase in $F$ until it reaches the steady-state equilibrium with the surroundings at $F = 10$ as depicted in Fig. 4e. From the progression of the temperature curve in the 2-D domain, it can be stated that the analytical solution produced by the thermal wave model is much pertinent to the actual model to forecast the thermal field compared to the classical model as it is a finite speed of the propagation of heat in absorber plates. Temperature plot at $F = 0.5$ is compared for $Ve = 0.5$ and $Ve = 0.0$ as shown in Figs. 4b and 4f, respectively to illustrate the influence of time-lag in the starting time span of solar heating. The fashion of temperature pattern is the same as the 2-D steady-state case presented in previous literature.[2] An analytical modeling in 2-D Cartesian coordinates was presented with the identical design situation. A straight-wise comparative evaluation is not applicable due to the dissimilarity in the boundary condition at $X = 1$. This investigation is particularly helpful for understanding a combined impact of heat conduction in x- and y- directions of the collector plate under the solar insolation in comparison to the 1-D model, which describes the heat transmission in only one direction.

Then the inspection done on the basis of the thermal wave model is to create the thermal response for different Vernotte numbers. 3-D thermal contours provide the idea for the heat transfer direction in the physical domain of an absorber plate. Fig. 5 demonstrates the 3-D surface isotherms for different Fourier and Vernotte numbers when the solar collector is receiving solar radiation ($S^* = 1$). Two-dimensional temperature field produced during heating of collector plate at $F = 0.5$ and $F = 2.5$ for $Ve = 1.0$ is shown in Figs. 5a and 5b, respectively. The thermal energy is increasing in the direction of the flow of collector-fluid. Obviously, there are a noteworthy impact on the energy propagation in the physical domain in both x- and y- directions for the variation of the entry energy of the collector fluid. Thermal contour lines are normal to the boundaries at $x = 0$ $(0 \leq y \leq L)$, $y = 0$, and $y = L$ $(0 \leq x \leq L)$. At the boundary adjacent to the collector fluid-carrying tube, i.e. $x = L$ $(0 \leq y \leq L)$, the contour lines incline at the boundary. By comparing the isotherm lines of Figs. 5a and 5b with Figs. 4b and 4d, respectively, it is observed that the overall plate temperature has been decreased with an increase in thermal relaxation time. In Figs. 5c and 5d illustrated for $Ve = 2.5$, the same thermal phenomena as shown in the previous figures are obtained associated with a lower magnitude of plate temperature. For $F = 0.5$, the wave nature of the temperature plot has once again been observed due to the high lag of thermal flux in the 2-D domain. Furthermore, the figure demonstrates a slight decrease in temperature along the x-direction and the temperature at the line of symmetry is greater than the plate temperature at any given X-coordinate. With a further increase in the Fourier number from $F = 2.5$, a decreasing temperature trend is observed along the x-direction in Fig. 6d. As all the temperature curves are perpendicular to the boundaries having zero thermal gradients, this examination further supports the exactness of the mathematical model and the MATLAB program.

*3.2. Temperature response of the solar collector under stagnation condition*

An effort is dedicated to evaluating the thermal aspect with Fourier numbers of an absorber plate under the stagnation condition having a lagging time. The temperature of the absorber plate decreases as it cools down from its steady-state condition to the surrounding ambient temperature. In Fig. 6a, results have been shown for the isothermal temperature contour plot at $F = 0.01$ for $Ve = 0.5$. As the absorber plate is at a steady-state at the initial condition, a uniform temperature curve is observed in the figure. However, with further increase in Fourier number, a substantial departure of thermal characteristics occurs with respect to Fig. 6b plotted for $F = 0.2$ and $Ve = 0.5$. Here, it may be noted that the temperature curves show non-uniform characteristics, and the overall thermal energy of the plate declines due to the heat loss to the surroundings. Once more, 2-D contours are plotted for $F = 1$ and $F = 2$ in Figs. 6c and 6d, respectively. Here, with increasing Fourier number it has been observed that the contours become more uniform in nature and the plate temperature decreases towards the ambient temperature. This progress also rationalizes the lagging feature of heat flux in the solar collector as discussed in Fig. 5. At a low value of $Ve$, the temperature pattern is invariant with the classical model and the nonexistence of $Ve$ transforms the thermal wave model into the classical model. A higher value of $Ve$ generates a more irregular thermal response but as the magnitude of $F$ increases, this characteristic gradually omits. The temperature pattern in the physical domain is to satisfy the design situation in the respective direction and therefore, this result may authenticate the accurateness of the present mathematical analysis and computer programming. In Fig. 6, constant temperature lines intersect the boundary surfaces normally according to boundary conditions taken. From all these figures, isotherms portray an individual behavior with declining temperature values and non-uniform peaks along the x- and y-directions. The creation of 3-D surface plots as shown in Fig. 7 brings out the indication of energy transfer in an actual area, and temperature patterns at different locations of the absorber plate make sure us the expectation of the temperature lines in dissimilar thermal relaxation time lags. In Figs. 7a and 7b, a thermal isotherm is generated for $Ve = 1$ and $Ve = 1.5$, respectively at a constant Fourier number $F = 0.5$. The non-uniform temperature field based on the non-Fourier model doesn't alter according to the discussion made in the previous figure (Fig. 6). The temperature in the x-y domain declines at a lesser rate with an increase in $Ve$. As the thermal wave model for heat transfer introduces a finite speed of propagation of heat in an absorber plate, it leads to lowering an amount of heat transfer and thus the magnitude of temperature becomes higher as depicted in Fig. 7b. Therefore, this occurrence actually provides the acceptability of the present study and the trend of temperature variation matches with the previous figures. Furthermore, it is to note that the collector plate domain satisfies the boundary conditions taken.

*3.4. Effect of Fourier number on the temperature at the center of the absorber plate*

The temperature in the symmetric module of an absorber plate dependent on the Fourier number is illustrated in Fig. 8 for both starting and closing conditions. Fig. 8a illustrates the temperature pattern determined by both 2-D classical and thermal wave models at $X = 0.5$ and $Y = 0.5$ for a constant value of $S^* = 1$. Considering Ve in this study to produce results were 0.5, 2.5, and 3.5. For the classical heat transfer, it is to observe that the thermal energy gradually increases until it reaches a stable condition. At $Ve = 0.5$, a difference in predictions by the classical and thermal wave models is noticed. Owing to the time lag in the nature of heat flux in the absorber plate, the magnitude of dimensionless temperature for the thermal wave mode of heat transfer is lesser than the classical mode where no lagging exists. This result properly rationalizes the significance of the thermal wave condition in the theoretical model. Furthermore, a substantial difference was obtained at higher $Ve$ values. The pattern of temperature increments initially with $F$ in a uniform way reaches a maximal condition declines rapidly and then increments again with the time to achieve equilibrium steady condition. This illustrates the wavy nature of the temperature pattern as a function of $F$ at higher $Ve$ values. For $Ve = 2.5$ temperature curve attains first a peak value at $F = 9.5$ signifying the lagging behavior in heat propagation, causing a big deviation from Fourier model values. Again at $Ve = 3.5$ temperature crest has been found at $F = 13.3$, which actually denotes the increase in attenuation time of peak temperature with an increase in thermal relaxation time. Therefore, $Ve$ significantly affects the magnitude of plate temperature at initial $F$ during the starting of the solar collector. The fluctuating wavy nature of temperature response gradually diminishes at higher $F$ values to achieve the thermal equilibrium. The oscillating temperature is dominant at lower $F$ and higher $Ve$ values and a sufficiently high $F$ value can be used to dampen the fluctuating nature of the thermal wave model to merge with the classical model. The time taken by the absorber plate to have a peak thermal energy for various $Ve$ is shown to obtain the plate temperature in practical applications in comparison to the classical model, which is with the infinite speed of thermal propagation in the absorber plate. In Fig. 8b, temperature patterns are depicted as a functional relationship with $F$ for different $Ve$ at $X = 0.5$ and $Y = 0.5$ when the absorber plate is under the stagnation condition. It can be observed that the temperature pattern diminishes until it cools down to the surrounding ambient temperature. For the Fourier-based model, the absorber plate cools down almost instantaneously in the absence of solar radiation due to the infinite propagation speed of the thermal wave. However, for $Ve = 0.5$ the dimensionless temperature becomes almost zero approximately at $F = 2.2$. This result indicates that the present model based on the thermal wave approach is more suitable to estimate the time taken by the absorber plate to cool down from its steady-state. The primary purpose behind such mathematical modeling is to make the present analysis more towards a practicable approach. Therefore, it is to be necessary to analyze the influence of $Ve$ on the thermal response in the solar collector at stagnation condition. A similar tendency of results is found with further increment in $Ve$ but a higher $F$ is required to reach the ambient condition due to the lagging nature of heat flux. For $Ve = 1$ and $Ve = 1.5$, the temperature curve attains to zero at $F = 6.3$ and $F = 10.6$, respectively. It has already been discussed in earlier results that the temperature pattern is spatially dependent on the x- and y-coordinates but the effect of the Fourier number remains consistent irrespective of its location. Nevertheless, from all the figures it can be concluded that a noteworthy dissimilarity between 1-D and 2-D classical and thermal wave models is evident. The workability of the absorber plate depends exclusively on the temperature variation in the plate. It can be demonstrated that a declining the thermal conductivity of the plate material results in the increase in plate temperature. The whole range of plate temperature can be amplified significantly by increasing the design variables such as solar flux parameter $S^*$, absorber plate parameter $Z_0$, and ambient temperature $T_a$. This results an increased losses and diminishes the collector plate efficiency. Based on the nature of isotherms with the thermo-physical variables, it demonstrates the essentiality of the present 2-D thermal wave modeling subjected to a practicable solar heating system of a solar collector.

## 4. Conclusions

Conclusions are summarized based on the results outlined in the present article as follows:

(a) A closed-form expression of temperature field for the classical and thermal wave heat transfer in an absorber plate has been developed using the separation of variables in the 2-D Cartesian domain. Furthermore, a convective boundary condition is implemented according to practical aspects of heat transfer that takes place at the plate region where a fluid-carrying tube is attached. Comparing temperature curves with the published research work,[40] a good agreement of the 2-D analytical results has been obtained for a case study.

(b) The accurateness of the current 2-D mathematical model has been justified by comparing with results based on numerical methods and it reveals that the temperature pattern at a given Fourier number is a function of both x-and y- coordinates. This result indicates the essentiality of two-dimensional modeling rather than the existing 1-D heat transfer models for predicting the temperature response in the absorber plate accurately.

(c) Two different analytical methods have been implemented for two different cases: firstly, during starting of a solar collector with zero initial condition (ambient temperature) and secondly, during stagnation condition with an initial condition based on the steady temperature. Both the approaches are compared with the published theoretical work.

(d) 2-D classical and thermal wave isotherms plotted at different times based on both the classical and thermal wave analyses effectively rationalizes the boundary conditions used for the absorber plate. The thermal behavior of the collector plate under the solar radiation is discussed with appropriate explanations and a valid approach. The thermal wave model is to be more accurate than the classical model as it incorporates the lagging nature of heat propagation. The 3-D surface plots illustrate the flow pattern of heat in the test domain at different Fourier and Vernotte numbers.

(e) The influence of the Fourier number on the thermal aspect in a module of the absorber plate has been presented in this paper. The thermal wave characteristics appear when the thermal relaxation time is moderately high. Finally, it can be demonstrated that the time taken by the solar collector to cool down completely depends significantly on $Ve$.

**Data availability statement**

Data that support the finding of this study are available upon request.

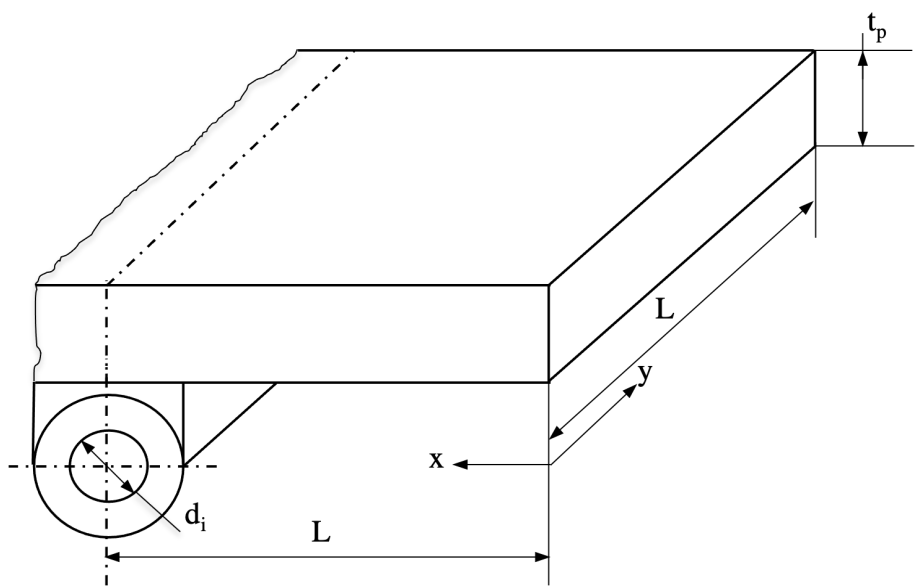

**Fig. 1.** A module for heat transfer of an absorber plate

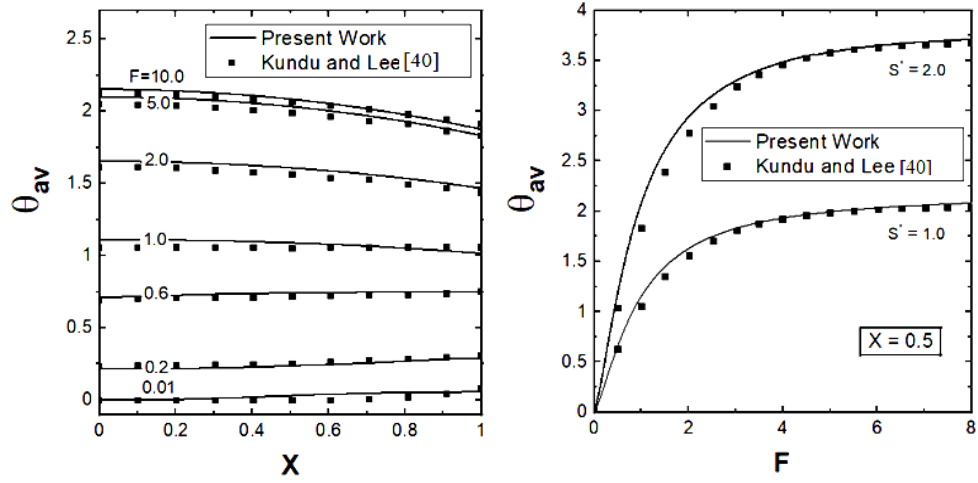

(a) $\theta_{av}$ vs. $X$ for different $F$   (b) $\theta_{av}$ vs. $F$ for different $S^*$ at $X$=0.5

**Fig. 2.** Comparison of average plate temperature as a function of $X$ for the present work with the published work [40] based on thermal wave analysis

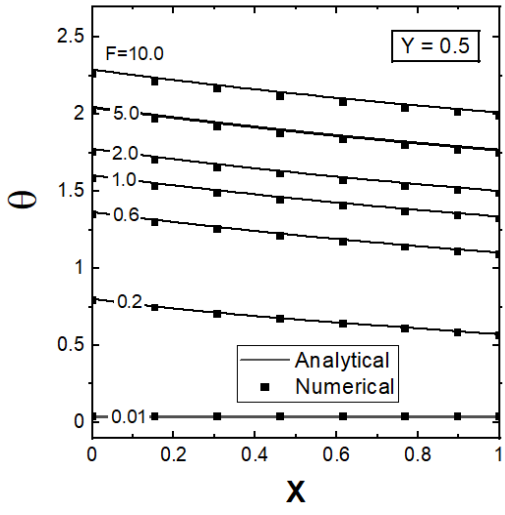
(a) Fourier, Y=0.5

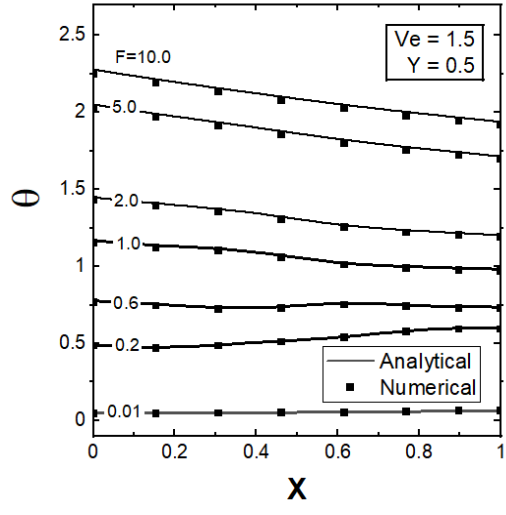
(b) Non-Fourier, Y=0.5

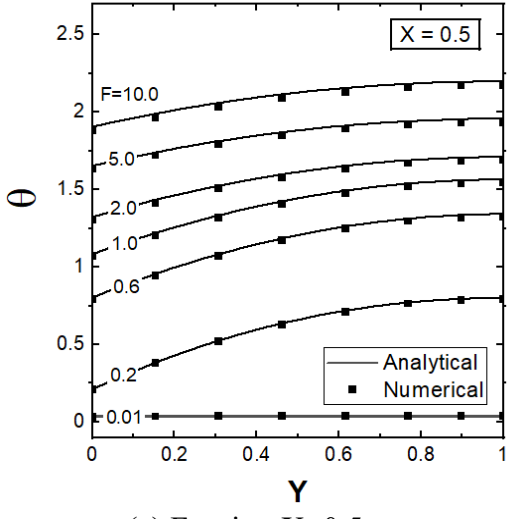
(c) Fourier, X=0.5

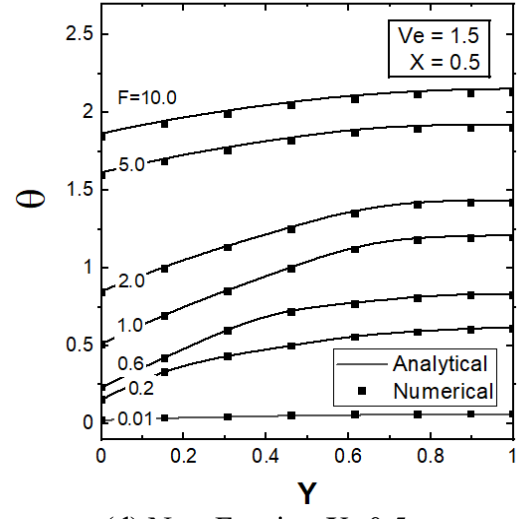
(d) Non-Fourier, X=0.5

**Fig. 3.** Classical and thermal wave temperature patterns in a heat transfer module for different $F$, calculated by numerical and analytical methods

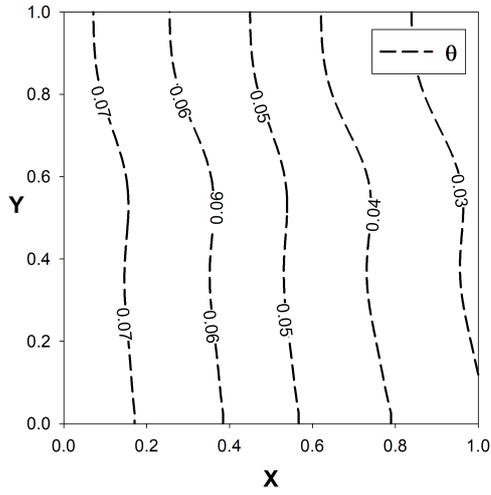
(a) $F = 0.01$, Ve=0.5

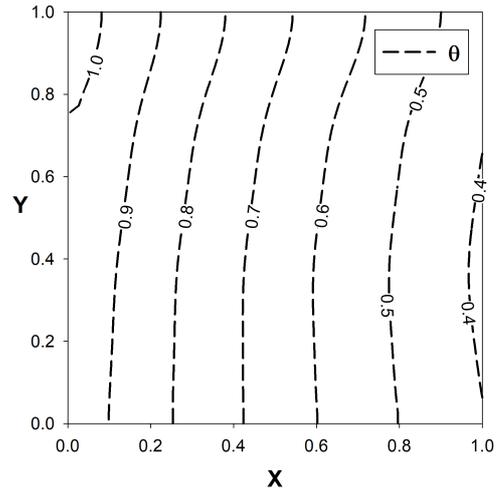
(b) $F = 0.5$, Ve=0.5

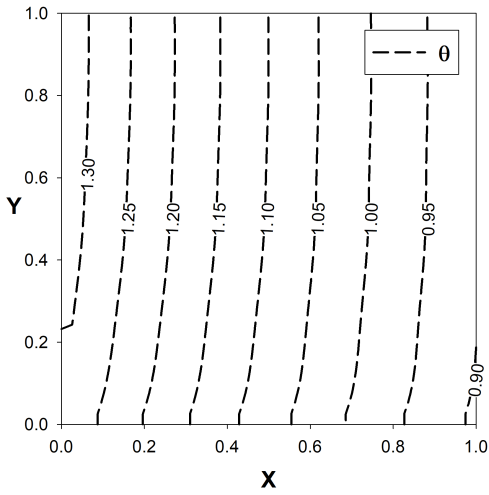
(c) $F = 1.0$, Ve=0.5

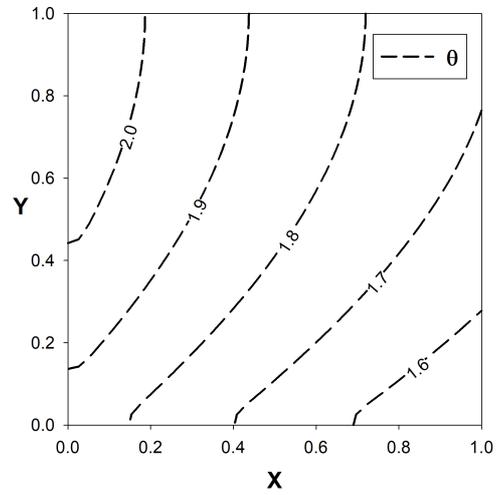
(d) $F = 2.5$, Ve=0.5

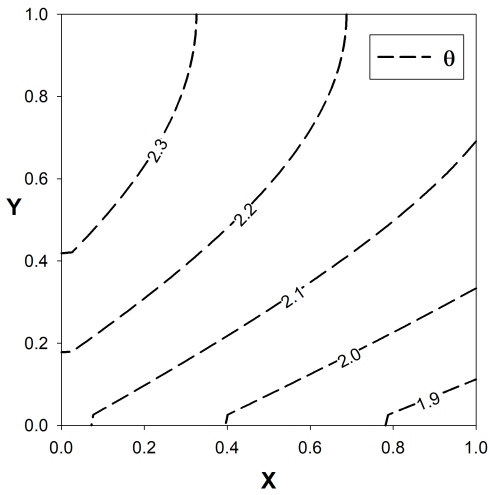
(e) $F = 10.0$, Ve=0.5

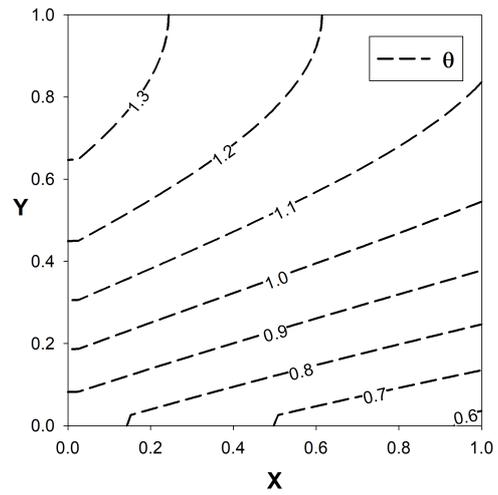
(f) $F = 0.5$, Ve=0

**Fig. 4.** 2-D thermal contour based on classical and thermal wave analyses at different Fourier numbers

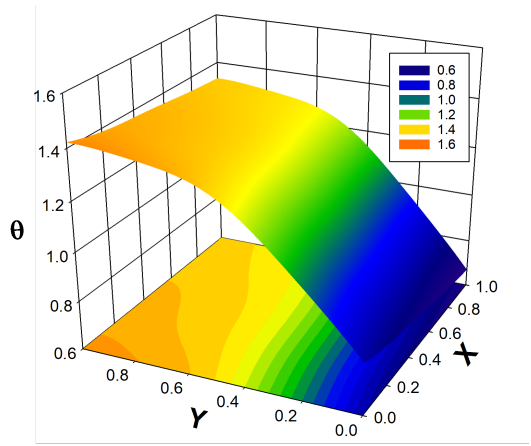

(a) $F = 0.5$ and $Ve = 1.0$

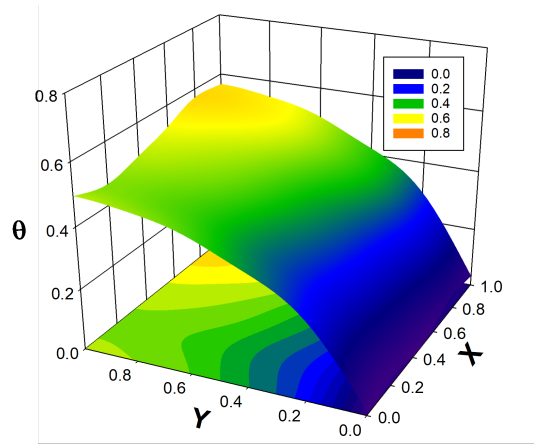

(b) $F = 0.5$ and $Ve = 2.5$

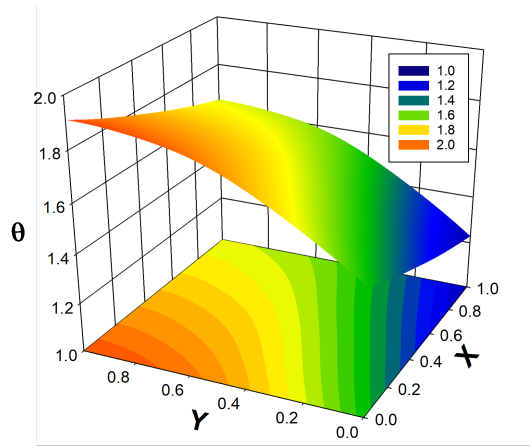

(c) $F = 2.5$ and $Ve = 1.0$

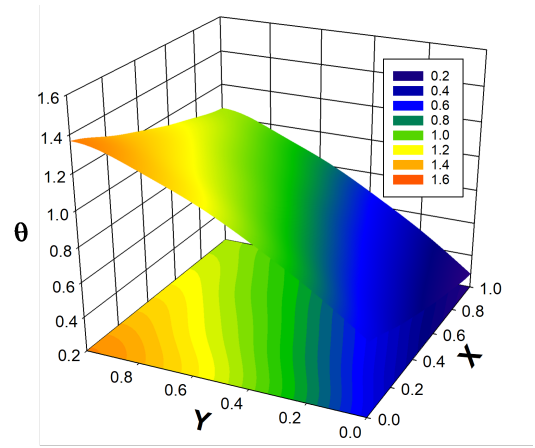

(d) $F = 2.5$ and $Ve = 2.5$

**Fig. 5.** 3-D thermal surface plots of $\theta$ at different space coordinates and at different $F$ and $Ve$

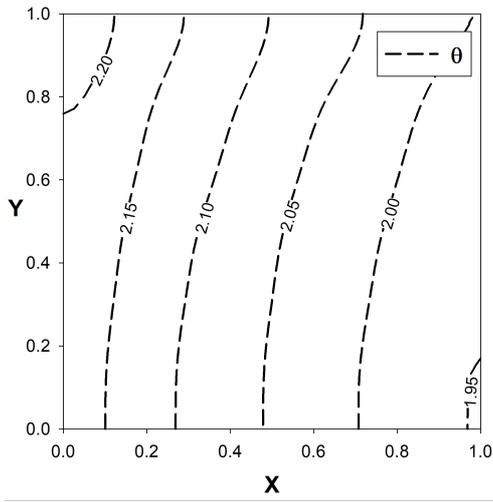
(a) $F = 0.01$ and Ve=0.5

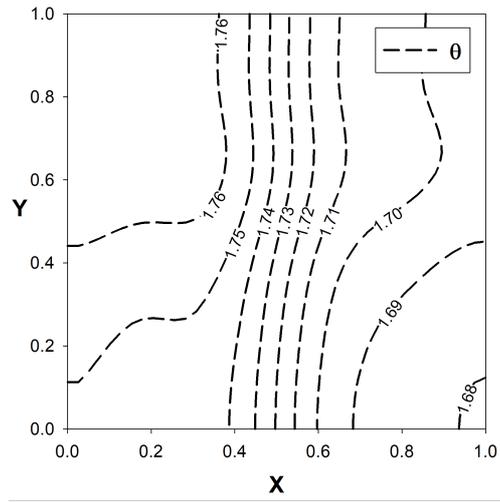
(b) $F = 0.2$ and Ve=0.5

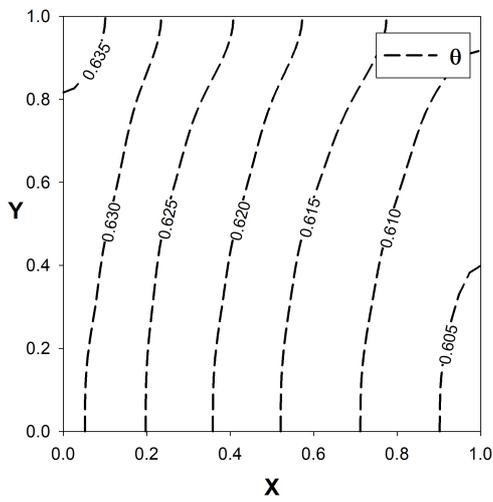
(c) $F = 1.0$ and Ve=0.5

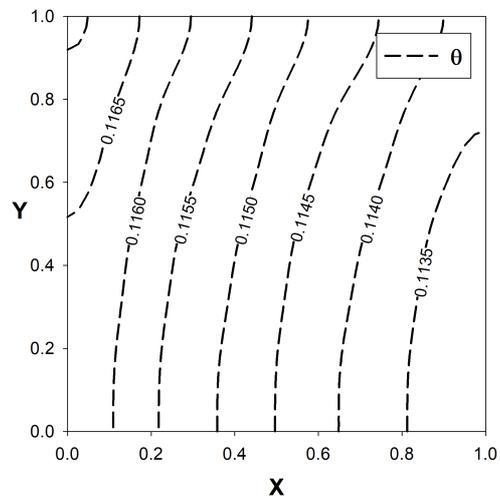
(d) $F = 2.0$ and Ve=0.5

**Fig. 6.** Isotherms for thermal wave analysis at $Ve = 0.5$ for different Fourier numbers under stagnation condition

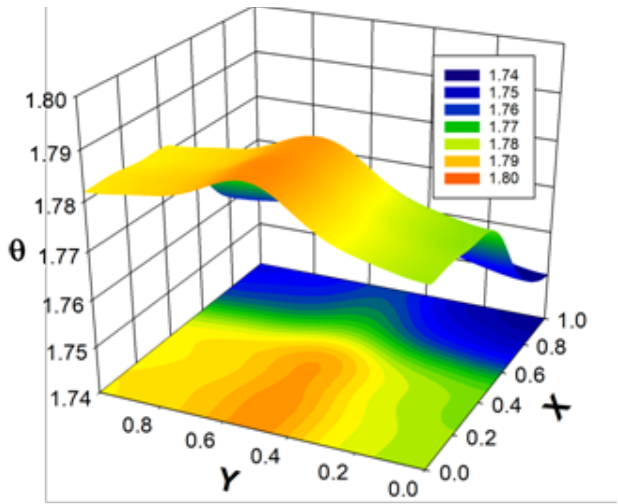
(a) $F = 0.5$ and $Ve = 1.0$

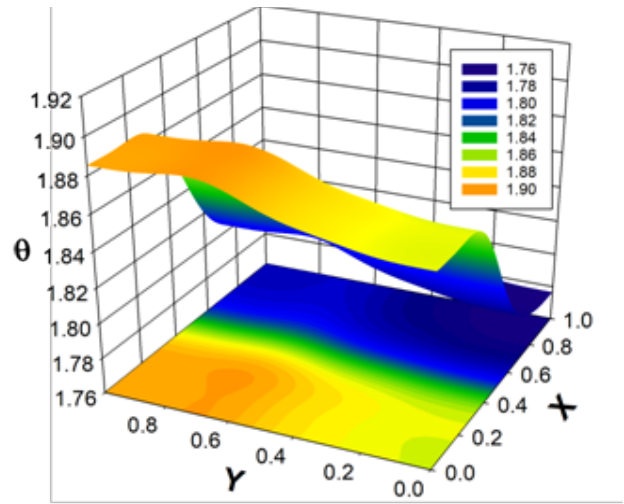
(b) $F = 0.5$ and $Ve = 1.5$

**Fig. 7.** Thermal contours of $\theta$ at different space coordinates and at different $F$ and $Ve$ under stagnation condition

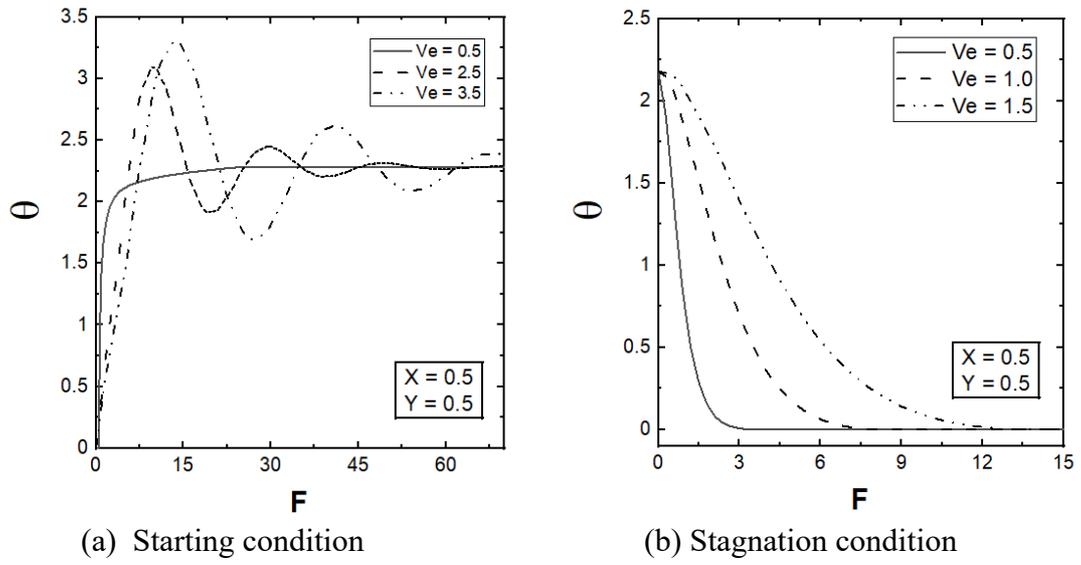

(a) Starting condition  (b) Stagnation condition

**Fig. 8.** Variation of *F* on the temperature field at the middle of a module for different Vernotte numbers